\documentclass[aps,twocolumn,amsmath,amssymb,showpacs,prb,floatfix]{revtex4}
\usepackage{ulem}
\usepackage{amsmath}
\usepackage{color}
\usepackage{graphicx}
\usepackage{epsfig}
\usepackage{SIunits}
\usepackage{mathtools}
\newlength{\upit}\upit=0.1truein

\newcommand{\ltappr}{{{\lower4pt\hbox{$<$} } \atop \widetilde{ \ \ \ }}}
\newlength{\bxwidth}\bxwidth=1.5 truein

\newcommand{\dg}{^{\dagger }}

\newcommand{\up}{\uparrow}
\newcommand{\dw}{\downarrow}

\newcommand{\hast}{{\Psi}}

\newlength{\figwidth}
\newlength{\shift}
\newlength{\fight}
\newcommand{\fg}[3]
{
\begin{figure}[ht]

\vspace*{-0cm}
\[
\includegraphics[width=\fight]{#1}
\]
\vskip -0.2cm
\caption{\label{#2}
\small #3
}
\end{figure}}

\newcommand \bea {\begin{eqnarray} }
\newcommand \eea {\end{eqnarray}}
\newcommand{\bk}{{\bf{k}}}

\newcommand{\bQ}{{\bf{Q}}}
\newcommand{\bR}{{\bf{R}}}
\newcommand{\urs}{URu$_{2}$Si$_{2}$\ }
\newcommand{\ursp}{URu$_{2}$Si$_{2}$}
\newcommand{\psic}{{c}}

\def\joinrelde{\mathrel{\mkern-11mu}}
\def\joinrel{\mathrel{\mkern-9mu}}

\def\joinrelw{\mathrel{\mkern-6mu}}

\def\relbd{\mathrel{{\bf\smash{{\phantom- \above1pt \phantom-
}}}}}
\def\ltdash{\raise-1.8pt\hbox{$\scriptscriptstyle |$}}
\newcommand{\contract}[1]{\stackrel{\mathclap{\displaystyle \ltdash
\joinrel\relbd
\joinrelw\relbd
\joinrelw\relbd
\joinrelw\relbd\joinrel
 \ltdash
\joinrelde\relbd
\joinrel\relbd
\joinrel\relbd
\joinrel\relbd
\joinrel\relbd\joinrel
 \ltdash
}}{#1}}
\newcommand{\contracty}[1]{\stackrel{\mathclap{\displaystyle
\ltdash
\joinrel\relbd
\joinrelw\relbd
\joinrelw\relbd
\joinrelw\relbd\joinrel
\ltdash
}}{#1}}

\begin{document}
\title{Ising Quasiparticles and Hidden Order in $URu_2Si_2$}
\author{Premala Chandra,$^1$ Piers Coleman,$^{1,3}$ and Rebecca Flint$^2$}
\affiliation{$^1$ Center for Materials Theory, Department of 
Physics and Astronomy, Rutgers University, 
Piscataway, NJ 08854 USA}
\affiliation{$^2$ 
Department of Physics and Astronomy, Iowa State University, 
12 Physics Hall, Ames, Iowa 50011 USA}
\affiliation{$^{3}$ Department of Physics, Royal Holloway, University
of London, Egham, Surrey TW20 0EX, UK.}
\date{\today}
\begin{abstract}
The observation of Ising quasiparticles is a signatory feature of the
hidden order phase of \ursp.  In this paper we discuss its nature 
and the strong constraints it places on current theories
of the hidden order.  In the hastatic theory such anisotropic
quasiparticles are naturally described described by resonant
scattering between half-integer spin conduction electrons
and integer-spin Ising moments. The hybridization that
mixes states of different Kramers parity is spinorial;
its role as an symmetry-breaking order parameter is consistent
with optical and tunnelling probes that indicate its sudden
development at the hidden order transition. 
We discuss the microscopic origin of hastatic order, identifying it 
as a fractionalization of three body bound-states into integer spin
fermions and half-integer spin bosons. 
After reviewing key features of hastatic order and their
broader implications, we discuss our predictions for experiment and
recent measurements.  We end with challenges both for hastatic order
and more generally for any theory of the hidden order state in \urs.
\end{abstract}
\maketitle
\section{Introduction}

We begin by noting that two key 
developments in heavy Fermion physics that relate
to the hidden order problem in \urs 
were both published in Philosophical Magazine. Forty years ago, Neville Mott\cite{Mott74}
pointed out that the development of coherence in heavy electron
systems should be understood as a hybridization of f-electrons
connected with the Kondo effect. Twenty five years later, 
Okhuni et al.\cite{Okhuni99} discovered that 
in the hidden order phase, the mobile carriers are Ising quasiparticles.
This paper discusses how
these two phenomena - the development of an emergent hybridization and
the formation of pure Ising quasiparticles are inextricably linked
with the hidden order in \urs.

There is still no consensus on the nature of the ``hidden order'' phase in \urs despite
several decades of active theoretical and experimental research.\cite{Palstra86,Schlabitz86,Mydosh11}  At $T_{HO}=17.5 K$ there are
sharp features in thermodynamic quantities and a sizable
ordering entropy ($ S > \frac{1}{3} R \ln 2$); however there is no observed 
charge order,
and spin ordering in the form of
antiferromagnetism occurs only at finite pressures.\cite{Palstra86,Schlabitz86,Mydosh11,Broholm91,Amitsuka07,Takagi07}
At first sight, it seems straightforward to link hidden order 
to the formation of a ``heavy density wave'' 
within a pre-formed  heavy electron fluid. Since 
there is no observed magnetic moment or charge density 
observed in the hidden order (HO) phase, such a
density wave must necessarily involve a higher order multipole
of the charge or spin degrees of freedom and various theories of this sort
have indeed been 
advanced.\cite{Niewenhuys87,Ramirez92,Walter93,Santini94,Kasuya97,Chandra02,Varma06,Elgazzar09,Balatsky09,Haule09,Harima10,Haule10,Oppeneer10,Dubi11,Pepin11,Fujimoto11,Kusinose11,Yuan12,Ikeda12,Riseborough12,Rau12,Ressouche12,Das14} 
In each of these scenarios, 
the heavy electrons develop coherence via a crossover at
higher temperatures, and the essential hidden order is then a
multipolar charge or spin density wave.  However such multipolar order can not naturally
account for the emergence of heavy Ising quasiparticles, a signature feature of \urs that has been
probed by two distinct experiments.\cite{Brison95,Okhuni99,Altarawneh11,Altarawneh12}
The essential point here is that conventional 
quasiparticles have half-integer spin and are magnetically isotropic; they thus lack the essential 
Ising protection required by observation. In addition optical and tunnelling probes\cite{Santander09,Schmidt10,Aynajian10,Park11,Nagel12}
indicate that the hybridization in \urs develops abruptly at $T_{HO}$ and is thus associated with a 
global broken symmetry;\cite{Dubi11,Yuan12,Chandra13,Flint14} this is to be contrasted with the usual situation in heavy fermion materials
where it is simply a crossover. 

\fight=\columnwidth
\fg{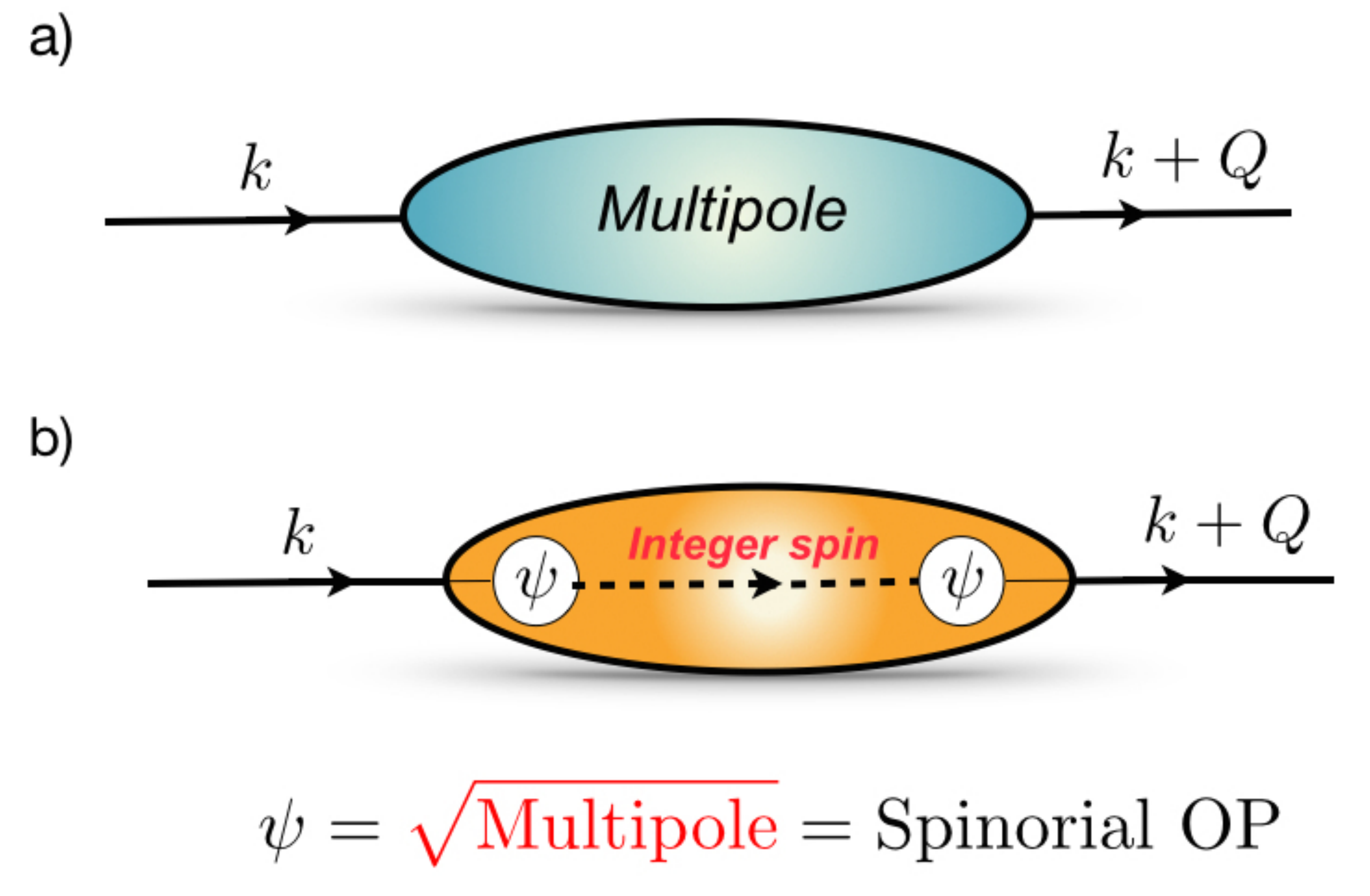}{Fsqrt}{Schematic contrasting the  multipolar and
spinorial theories of Hidden order. (a) in a multipolar scenario, the
heavy electrons Bragg diffract off a staggered spin or charge
multipole (b) in the hastatic scenario, the development of a spinor
hybridization opens up resonant scattering with a an integer spin
state of the ion. The multipole is generated as a consequence of two
spinorial scattering events. In this way, the Hastatic spinor order
parameter can be loosely regarded as the square root of a multipole. }

Here we argue that the elusive nature of the ``hidden order''  in \urs is 
{\sl not} due to 
its intrinsic complexity but rather that it results from a fundamentally
new type of order parameter.
In the ``hastatic'' proposal\cite{Chandra13,Flint14},
the observation of heavy Ising 
quasiparticles\cite{Brison95,Okhuni99,Altarawneh11,Altarawneh12} 
suggests
resonant scattering between half-integer spin
electrons and integer spin local moments, and the development of an 
spinorial order parameter.  It is perhaps useful to
contrast the various staggered multipolar scenarios for the hidden order with
the hastatic one proposed here. 
In the former, mobile f-electrons Bragg diffract off a multipolar
density wave (see Fig \ref{Fsqrt} (a)), 
whereas in the latter, the multipole contains an internal structure,
associated with the resonant scattering into an integer spin f-state 
(Fig \ref{Fsqrt} (b)).   
Hastatic order can thus be loosely regarded as the ``square root'' of a 
multipole order parameter;
in other words we argue that the origin of hidden order is
not a complex multipole but instead is an elementary
"half-tu-pole" that mediates hybridization between an Ising non-Kramers
doublet and the mobile conduction electrons.


Because the observed magnetic anisotropy of the heavy quasiparticles
is central to our approach, we'll begin by discussing these 
experiments\cite{Brison95,Okhuni99,Altarawneh11,Altarawneh12} 
in detail.  Next we'll review ``highlights'' of the 
hastatic proposals\cite{Chandra13,Flint14}, and
the broader implications of an order parameter that transforms
under double-group ($S=\frac{1}{2}$) representations.  Experimental
predictions and recent measurements will be discussed next.  We'll
end with challenges both for hastatic order and more generally
for any theory of hidden order in \urs.

\fg{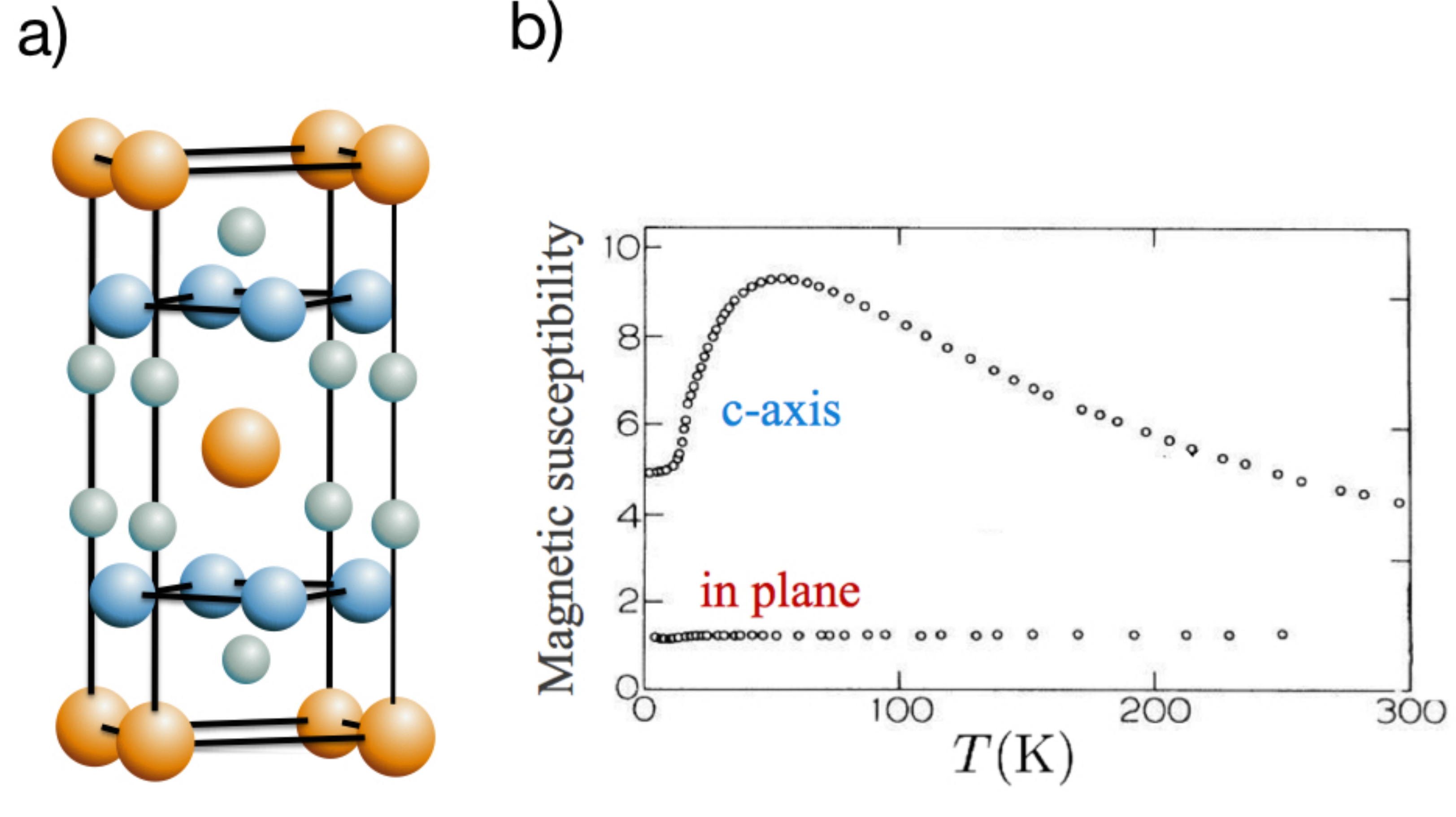}{fig2}{(a) Body-centered tetragonal 
structure of \urs (b) Measured anisotropic temperature-dependent bulk 
magnetic susceptibility\cite{Palstra86} of \urs}

\section{Ising Quasiparticles} 

Remarkably Fermi surface 
magnetization experiments in the HO state of \urs indicate 
near-perfect Ising anistropy in the g-factor ($g(\theta)$) of the
quasiparticles.\cite{Okhuni99,Altarawneh11}  
Measurements of the bulk susceptibility of \urs 
do show 
a strong Ising anisotropy along the c-axis (see Fig. \ref{fig2});\cite{Palstra86,Schlabitz86,Ramirez92,Mydosh11} 
this feature persists in dilute 
samples (U$_x$Th$
_{1-x}$Ru$_2$Si$_2$ with $x \sim 0.07$) 
suggesting that it is a single-ion effect.\cite{Amitsuka94}. However,
the Ising anisotropy of the bulk susceptibility is about a factor of
five, whereas the anisotropy in the Pauli susceptibility of the heavy
Fermi surface in the hidden order phase is in excess of 900.

According to  Onsager's  treatment of
a Fermi surface, the Bohr-Sommerfeld quantization of
quasiparticle orbits leads to a quantization of the area in k-space
according to  $\oint  dk_{x}dk_{y }= A (\epsilon_{n}) 
= (n+\gamma) \left(\frac{(2\pi)^{2}eB}{h}
\right)$ where $\gamma$ is a constant Berry phase term and $\epsilon$
is the Kinetic energy  of the Bloch waves (i.e energy without Zeeman
splitting)\cite{Schoenberg84}.
This condition 
leads to quantized kinetic energy $\epsilon_{n}= \hbar \omega_{c}
(n+\gamma)$.
When the Zeeman spin spitting is included, one finds that 
the quantized energies are given by\cite{Holtham73,Lonzarich73}
\begin{equation}\label{quantizedlevels}
E_{n\pm } = 
\overbrace {(n + \gamma)\hbar  \omega_{c}}^{\epsilon_{n}} \mp \frac{1}{2}
{g\mu_{B}B}, 
\end{equation}
where $\omega_{c} =
\frac{eB}{m^{*}}$ is the cyclotron frequency,
\begin{equation}\label{}
{m^{*}} = \frac{\hbar ^{2}}
{2 \pi}\frac{\partial A}{\partial\epsilon}, 
\end{equation}
is the effective mass and 
\begin{equation}\label{}
g = \frac{\oint \frac{dk_{\perp }}{v_{F}} g (\bk)}
{\oint \frac{dk_{\perp }}{v_{F}}}
\end{equation}
is the average of the g-factor over the orbit.  
Notice that the Onsager quantization condition means that the
kinetic energies of the up and down Fermi
surfaces are identical with the Zeeman splitting superimposed. 

The discrete 
summation over these quantized 
energy levels gives rise to an oscillatory component
in the magnetization given by \cite{Schoenberg84}
\begin{equation}
M \propto \sum_{\pm }
\sin\left(\frac{2\pi \mu_{\pm}}{\hbar \omega_{c}}\right)  
= \sum_{\sigma }\sin\left[
\frac{2\pi \mu}{\hbar \omega_{c}}\pm 2 \pi
\left( 
\frac{\frac{g}{2}\mu_B B}{\hbar\omega_{c} }
\right) 
 \right],
\end{equation}
where  $\mu_{\sigma } = \mu + \frac{\sigma }{2}g\mu_BB$ is the
Zeeman-split chemical potential. 
Summing the two terms together 
\begin{equation}\label{}
M \propto 2 \sin\left(\frac{2\pi \mu}{\hbar \omega_{c}}\right)  
\cos \delta
\end{equation}
where
\begin{equation}\label{}
\delta = 2 \pi\left(\frac{g\mu_B B}{\hbar  \omega_{c}} \right) =  \pi
\left(\frac{m^{*}}{m}
 \right)
\end{equation}
is the phase shift induced by the Zeeman splitting. Notice that
$\delta $ is field indendent, so it affects the overall amplitude
without changing the dHvA frequencies. 
In particular in systems where the g-factor is a strong function of angle,
namely in orbits where the Zeeman splitting is a half-integer multiple of the
cyclotron energy, the up and down Fermi surfaces destructively
interfere to produce a ``spin zero''; here the dHvA signal
identically vanishes and 
\begin{equation}
\alpha_{n}=2 \pi\frac{\hbox{Zeeman splitting}}{\hbox{cyclotron energy}} = 
g(\theta_{n}) \frac{m^*}{2m_e} = n + \frac{1}{2}
\end{equation}
where $n$ is a positive integer and $\theta_n$ is the (indexed) angle
with respect to the c-axis.  The observation of spin zeroes in dHvA
thus provides a way of detecting the presence of a spin-degenerate
Fermi surface and, provided the indexing can be done reliably, enables
a direct measurement of the dependence of the g-factor $g (\theta )$ on the
orientation of the orbit.

\fg{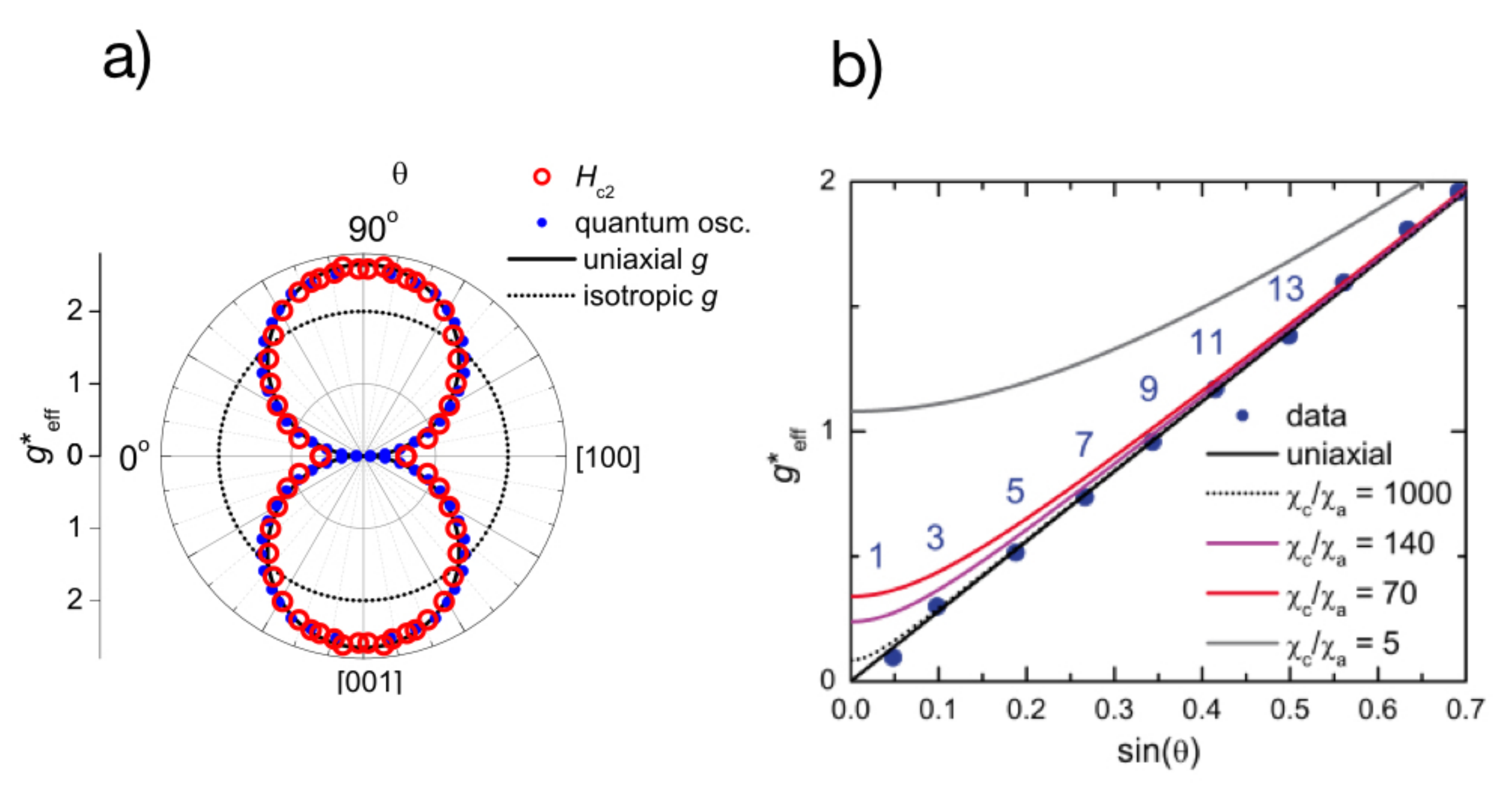}{fig3}{ Anisotropy of the g-factor of quasiparticles in 
\urs (a) plotted in polar coordinates derived from spin zeroes in quantum oscillation measurements and the anisotropy of the upper critical field (b) versus sine of the angle out of the basal plane, showing that the data requires a Pauli susceptbility anisotropy in excess of 900.\cite{Brison95,Okhuni99,Altarawneh11,Altarawneh12}}

Sixteen such spin zeroes are observed (cf. Fig. 3) in the HO state of URS,\cite{Okhuni99,Altarawneh11} 
in contrast to the one per band seen
in the cuprates.\cite{Ramshaw10} At the most elementary level, these results
tell us that the heavy $\alpha $ pocket of the HO state involves
quasiparticles that carry spin, with a two-fold degeneracy at
each point in k-space. It is well known that such degeneracies survive 
strong spin-orbit coupling if there is inversion symmetry combined with 
time-reversal invariance or a combination of time-reversal and translational
invariance as in a commensurate spin density wave.
Moreover, we can place stringent bounds on the level of perfection 
of both the degeneracy and the Ising anisotropy. 

The Zeeman splitting scales from  more than fifteeen times the cyclotron
frequency along the c-axis to less than half a cyclotron frequency
along the basal plane.  This puts a rigorous bound on the g-factor
anisotropy
\begin{equation}
\frac{g_\perp}{g_c} < \frac{1}{30}
\end{equation}
where $g_\perp = g(\theta_n \sim \frac{\pi}{2})$ and $g_c = (\theta_n = 0)$, 
indicating that 
the splitting energy between the orbits depends {\sl only} on the c-axis 
component of the applied magnetic field ($B_c$), namely that
\begin{equation}
g(\theta_n) = g^* \cos \theta_n
\end{equation}
where $g^* = 2.6$ in contrast to the 
isotropic $g=2$ for free electrons.\cite{Okhuni99,Altarawneh11}   
We note that these dHvA oscillations were generated by the heavy $\alpha$
pockets of \urs, and thus could be argued to come from a select region
of its Fermi surface.
However this magnetic anistropy
is also observed in the angular dependence of the upper critical
field $H_{c2}(\theta)$ that is sensitive to the entire heavy fermion
pair condensate.\cite{Brison95,Altarawneh12} 
The $g(\theta)$ derived from $H_{c2}(\theta)$ matches
that from the dHvA measurements very well for angles near the c-axis where
$H_{c2}$ is Pauli-limited.\cite{Altarawneh12}  
However the anisotropic bound on the g-factor is 
less stringent than that found from the quantum oscillation experiments,
since the in-plane $H_{c2}$ is larger than expected, probably due to
orbital contributions.  Returning to the bounds placed
by the spin-zeroes measurements, we note that since 
the Pauli susceptibility $\chi^P$ 
scales with the {\sl square} of 
the g-factor, these resolution-limited measurements of $\frac{g_c}{g_\perp}$ 
suggest that 
\begin{equation}
\chi^P (\theta) = \chi^{P*}\cos^2 \theta \qquad
\frac{\chi^P_c}{\chi^P_\perp} > 900.
\end{equation}
Such a large anisotropy should be directly observable in electron spin 
resonance measurements that probe the Pauli susceptibility directly
in contrast to bulk susceptibility measurements where  
Van Vleck contributions are also present.

To our knowledge, this is the largest number of spin zeroes that
have ever been observed in any material; furthermore the Ising nature
of the quasiparticles in the hidden order state is a dramatic
departure from the usual magnetic isotropy of free conduction electrons.
A natural explanation for the quasiparticle Ising anisotropy is 
that the Ising character of the uranium ions has been 
transferred to the quasiparticles via hybridization, and this is a 
key element of the hastatic proposal.\cite{Chandra13,Flint14}
The giant anisotropy in $\frac{g_\perp}{g_c}$, places a strong constraint 
on the energy-splitting $\Delta$ between the two Ising states. This
quantity must be smaller than half a cyclotron frequency, or 
\begin{equation}
\Delta < \frac{1}{2}\hbar \omega_{c}.
\end{equation}
In the dHvA measurements, the effective mass on the $\alpha $ orbits 
is $m^{*}= 13 m_{e}$, and the measurements were made at $B=13T$, giving
\begin{equation}\label{}
\frac{\Delta }{k_{B}}
\ltappr \left(
\frac{\hbar e B}{2 (m*/m_{e})m_{e}}
\right)
= 0.67 {\rm K}.\end{equation}
Additional support for a very small $\Delta$ comes from the dilute 
limit,\cite{Amitsuka94}
U$_x$Th$_{1-x}$Ru$_2$Si$_2$ (x = .07), 
where the Curie-like single-ion behavior
crosses over to a critical logarithmic temperature dependence below $10K$,
$\log T/T_K$, where $T_K \approx 10K$. 
This physics has been attributed to two-channel Kondo 
criticality, again requiring a splitting $\Delta \ll 10K$.

Constrained by the anisotropic bulk spin susceptibility and the 
quantum spin zeroes, we therefore require the U ion to be an Ising 
doublet with the form
\begin{equation}
|\Gamma_\pm \rangle  = \sum_n a_n |\pm (J_z - 4n)\rangle,
\end{equation}
where the addition and subtraction of angular momentum in units of
$4\hbar $ is a consequence of the four-fold symmetry of the \urs tetragonal
crystal. However, the presence of a perfect Ising anisotropy requires an 
{\sl Ising selection rule}
\begin{equation}\label{sum}
\langle \Gamma_\pm |J_\pm
|\Gamma_\mp\rangle  = 0
\end{equation}
that, in the absence of fine-tuning of the coefficients $a_n$, 
leads to the condition that $-(J_{z}+4n')\neq (J_{z}+4n)\pm 1$, or
$J_{z}\neq 2(n-n')\pm \frac{1}{2}$, requiring $J_z \in \mathbb{Z}$
{\sl must be an integer}.
For any generic half-integer $J_z$, corresponding to a Kramers doublet, the
selection rule is absent so that crystal fields mix the $J_z$ states
leading to isotropic magnetic properties.  Within the five-parameter
crystal-field Hamiltonian of \urs, a simulated annealling search yielded just
one finely tuned $5f^3$ (Kramers) state with nearly zero transverse
moment, but the fit to single-ion bulk properties was poor.\cite{Flint12}  
In the tetragonal crystalline environment of \urs, such Ising anistropy is 
most natural in a $5f^2$ ($J=4$) configuration of the uranium 
ion, but doublets with
integer $J$ in general do not enjoy the symmetry protection of their
half-integer (Kramers) counterparts. However in \urs a combination
of tetragonal and time-reversal symmetries protects a non-Kramers doublet
\begin{equation}
|\Gamma_5 \pm> = \alpha |J_z = \pm 3> + \beta |J_z = \mp 1>
\end{equation}
that is quadrupolar in the basal plane and magnetic along the
c-axis, and it has been proposed as the origin of the magnetic
anisotropy in both the dilute and the dense \urs;\cite{Amitsuka94,Chandra13,Flint14} 
this can
be checked with a direct benchtop test.\cite{Flint12}
In the hastatic proposal the Ising anisotropy of  the U $5f^2$ ions
is transferred to the quasiparticles via hybridization between integer $J$ 
local moments and half-integer $J$ conduction electrons, and this
mixing of Kramers parity ($K = (-1)^{2J}$) has important symmetry 
implications.\cite{Chandra13,Flint14}

\fg{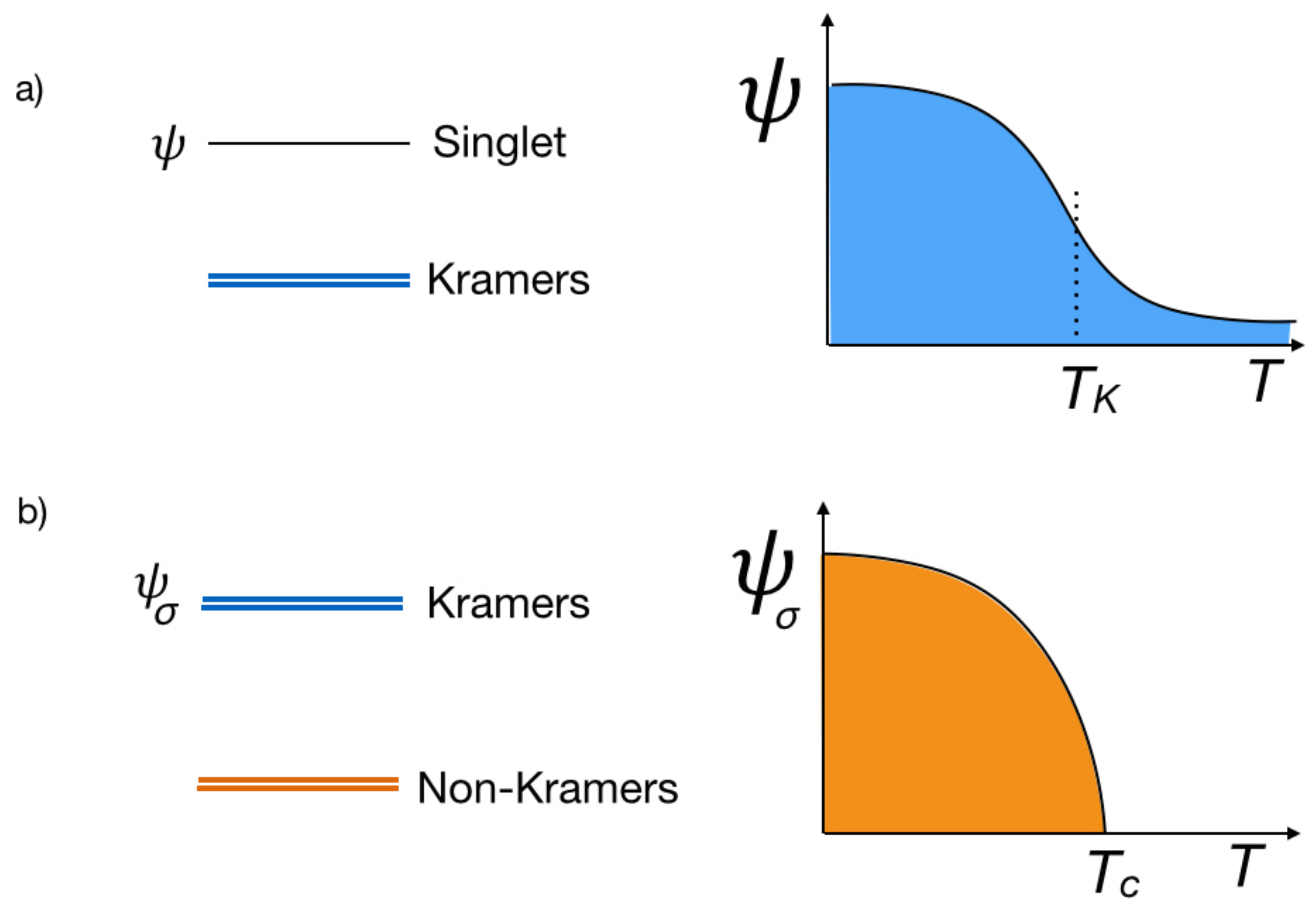}{fig4}{Schematic of (a) conventional (scalar) vs  (b)
spinorial hybridization where the hybridization is a) a crossover
and b) breaks spin-rotatinonal and time-reversal symmetries and thus
develops discontinuously as a phase transition.}

Conventionally in heavy fermion materials, hybridization involves
valence fluctuations between a ground-state Kramers doublet and an
excited singlet (cf. Fig. 4); in this case, hybridization is a scalar
that develops via a crossover leading to mobile
heavy quasiparticles.  However if the ground-state is a non-Kramers
doublet, the Kondo effect will involve an excited Kramers doublet (cf Fig. 4).
The quasiparticle
hybridization now carries a global spin quantum number
and has two 
distinct amplitudes that form
a spinor defining the hastatic order parameter
\begin{equation}
\Psi  = \left(\begin{matrix}
\psi_\up\cr
\psi_\dw\end{matrix}
\right).
\end{equation}
The onset of hybridization must break spin rotational invariance in addition
to single- and double time-reversal invariances via a phase transition; we 
note that optical, spectroscopic and tunneling probes\cite{Santander09,Schmidt10,Aynajian10,Park11,Nagel12} in \urs indicate the hybridization
occurs abruptly at the hidden order transition in contrast to the crossover
behavior observed in other heavy fermion systems (cf. Fig. 4). 

\section{Hastatic Order "Highlights"}

We next summarize the main points of the hastatic proposal,\cite{Chandra13,Flint14} noting that
the interested reader can find further discussion with more details elsewhere. 
Hastatic order captures the key features of the observed pressure-induced 
first-order phase transition in \urs between the hidden order and
and the  Ising antiferromagnetic (AFM) phases.\cite{Amitsuka07,Jo07,Villaume08,Hassinger10,Kanchavavatee11}
The most general Landau functional for the free energy density of a
hastatic state with a spinorial order parameter $\Psi $ as a function
of pressure and temperature is 
\begin{equation}\label{}
f[\Psi] = 
[\alpha (T_{c}-T)\vert \Psi \vert^{2}+ \beta \vert \Psi \vert^{4} - 
\gamma (\Psi \dg \sigma_{z}\Psi )^{2}
\end{equation}
and $\gamma = \delta(P-P_{c})$ where $P$ is pressure 
and the term $\gamma (\Psi \dg \sigma_{z}\Psi )^{2}$ determines
whether the direction of the spinor, either along the c-axis or
in the basal plane (cf. Fig. 5a). 

\fg{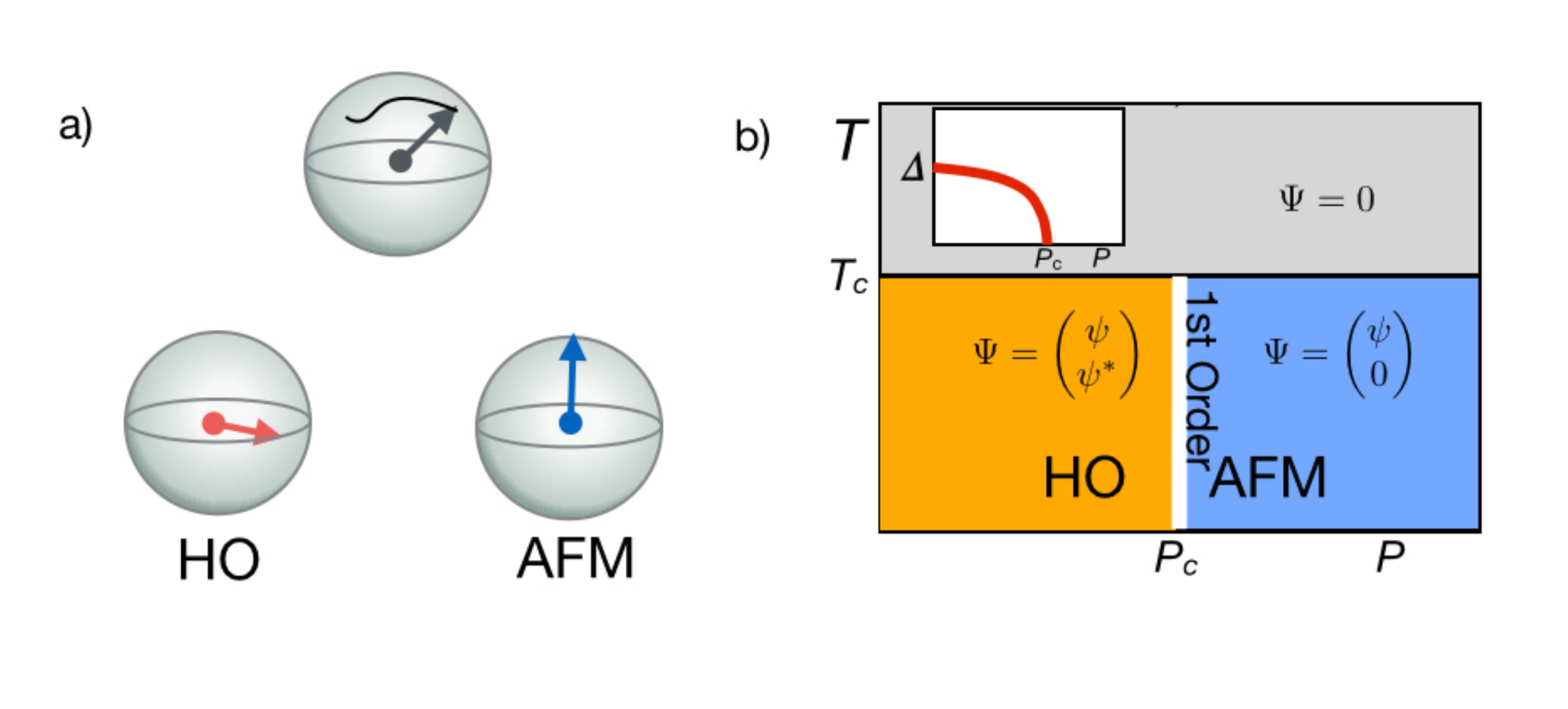}{fig5}{(a) The hastatic (hybridization) spinor
 disordered (at high tempertures) and ordered along the c-axis 
(Antiferromagnet) and in the basal plane (hidden order) (b) 
Temperature-Pressure Phase Diagram and the pressure-dependence of
the gap to longitudinal predicted by the hastatic theory}
 
Experimentally the $T_{AFM}(P)$ line is almost vertical, indicating
by the Clausius-Clapeyron relation that there is negligible
change in entropy between the HO and the AFM states.  Indeed
these two phases share a number of key features, including
common Fermi surface pockets; this has prompted the proposal that
they are linked by ``adiabatic continuity'', associated by
a notational rotation in the space of internal parameters.\cite{Haule10,Jo07}
This is easily accomodated with a spinor order parameter; for
the AFM phase ($P > P_c$), there is a large staggered Ising 
f-moment with
\begin{equation}
\Psi_A  \propto \left(\begin{matrix}
1\cr
0\end{matrix}
\right), \qquad 
\Psi_B  \propto \left(\begin{matrix}
0\cr
1\end{matrix}
\right)
\end{equation}    
corresponding to time-reversed spin configurations on alternating
layers $A$ and $B$.  For the HO state ($P < P_c$), the spinor
points in the basal plane 
\begin{equation}
\Psi_A  \propto \frac{1}{\sqrt{2}}\left(\begin{matrix}
e^{-i\phi /2}\cr
e^{i\phi /2}\cr\end{matrix}
\right), \qquad 
\Psi_B  \propto \frac{1}{\sqrt{2}}\left(\begin{matrix}
-e^{-i\phi /2}\cr
e^{i\phi /2}\cr\end{matrix}
\right)
\end{equation}    
and there is no Ising f-moment, consistent
with experiment, but Ising fluctuations do exist.
From this perspective the transition from HO to
AFM corresponds to a spin-flop of the two-component
hybridization order parameter from the basal
plane to the c-axis, and the resulting temperature-pressure
phase diagram is displayed in Fig. 5.  Generalizing this Landau
theory to study soft modes of the hastatic order,
we find that even though the transition at $P = P_c$
is first-order
the gap for longitudinal spin fluctuations decreases
continuously as
\[
\Delta \propto  |\Psi_{0} | \sqrt{P_{c}-P}.
\]
Since $dP_c/dT_c$ is finite, close to the transition, $\sqrt{P_c-P} \approx \sqrt{dP_c/dT_c(T-T_c)}$, and 
$\Delta \propto \sqrt{T-T_c}$.  Inelastic neutron scattering experiments can measure this gap 
as function of temperature at a fixed pressure where 
there is a finite-temperature first order transition, but to our knowledge a detailed study of this gap behavior has not yet been 
performed.  The iron-doped compound, URu$_{2-x}$Fe$_x$Si$_2$ can provide an attractive alternative 
to hydrostatic pressure, as iron doping acts as uniform chemical pressure and tunes the hidden 
order state into the antiferromagnet\cite{Kanchanavavatee11}.  The Landau theory can also be generalized to include
coupling to an applied magnetic field $B$, predominantly to $B_z = B \cos \theta$ due to the Ising
nature of the non-Kramers doublet; this then leads to an explanation of the observed large c-axis nonlinear susceptibility\cite{Ramirez92,Miyako91}
anomaly ($\Delta \chi_3$) in \urs, and a prediction of a 
large $\Delta \chi_3$ anisotropy, $\chi_3 \propto \cos^4 \theta$ where $\theta$ is the angle 
from the c-axis and the coupling coefficient must be determined from a microscropic approach.\cite{Chandra13,Chandra13b}

We use a two-channel Anderson lattice model to link hastatic order to the valence fluctuation physics
of non-Kramers doublets in \urs.  The $5f^2$ Ising $\Gamma_5$ 
ground-state configuration\cite{Amitsuka94} fluctuates to an excited $5f^3$ or $5f^1$ state via 
valence fluctuations.  The lowest lying excited state is most likely the $5f^3$ ($J=9/2$) state, but for 
simplicity we take it to be the symmetry equivalent $5f^1$ state, and assume that fluctuations 
to the $5f^3$ are suppressed; in this sense, we take an infinite-U two-channel Anderson model.
$\Gamma_7^+$ is taken to be the lowest energy doublet of the $5f^1$ state, and then the form of
the valence fluctuation Hamiltonian is determined by the orbital structure of the $\Gamma_5$ doublet.
Valence fluctuations occur in two orthogonal conduction electron channels, $\Gamma_7^-$ and $\Gamma_6$, and we find 
\begin{eqnarray}\label{l} H_{VF}(j) 
&=& V_6 \psic_{\Gamma_6 \pm}\dg(j) |\Gamma_7^+
\pm\rangle \langle \Gamma_5 \pm| \cr &+& V_7 \psic_{\Gamma_7 \mp}\dg(j)
|\Gamma_7^+ \mp \rangle\langle \Gamma_5 \pm| + \mathrm{H.c.}.
\end{eqnarray}
where $\pm$ denotes the ``up'' and ``down''
states of the coupled Kramers and non-Kramers doublets. The field
$\psic\dg_{\Gamma \sigma}(j) = \sum_\bk
\left[\Phi\dg_\Gamma(\bk)\right]_{\sigma \tau} c\dg_{\bk\tau}
\mathrm{e}^{-i\bk\cdot \bR_j}$ creates a conduction electron at
uranium site
$j$ with spin $\sigma$, in a Wannier orbital with symmetry $\Gamma\in
\{6,7\}$, while $V_6$ and $V_7$ are the corresponding hybridization
strengths. 
The full model is then written 
\begin{equation}\label{}
H = \sum_{ \bk \sigma }\epsilon_{\bk }c\dg_{\bk \sigma }c_{\bk \sigma}
+ \sum_{j}\left[H_{VF} (j)+ H_{a} (j) \right]
\end{equation}
where  $H_a(j) = \Delta E \sum_\pm|\Gamma_7 \pm,j\rangle \langle \Gamma_7 \pm,j| $
is the atomic Hamiltonian.

\fg{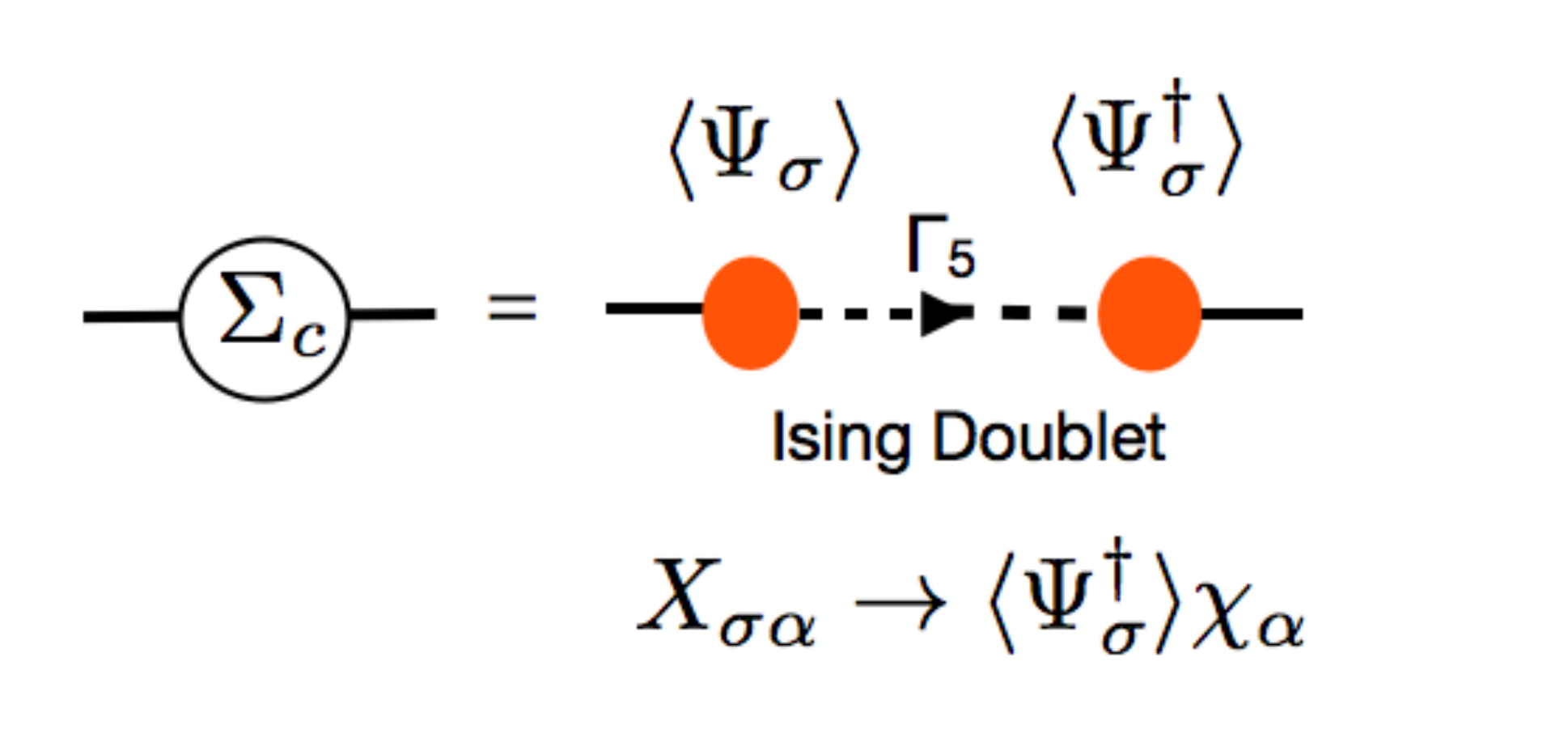}{fig4}{The conduction electron self-energy
$\Sigma_{c}$. 
Hybridization with spinorial order
parameter $\langle  \Psi_{\sigma }\rangle$ permits the development of
a $\Gamma_{5}$ Ising resonance inside the conduction sea, 
represented by the above Feynman diagram. 
}

Hastatic order is revealed by factorizing the Hubbard operators
\begin{equation}\label{}
X_{\sigma \alpha }= |\Gamma_7^+ \sigma\rangle\langle \Gamma_5 \alpha| = \hat \hast\dg_\sigma \chi _\alpha.
\end{equation}
Here $|\Gamma_5 \alpha\rangle = \chi\dg_\alpha
|\Omega\rangle$ is the non-Kramers doublet, represented by the
pseudo-fermions
$\chi \dg_{\alpha }$, while
$\hat \hast_\sigma\dg$ are {slave} bosons{\cite{Coleman83}} representing
the excited $f^{1}$ doublet
$|\Gamma_7^+ \sigma\rangle
=\hat\hast\dg_\sigma|\Omega\rangle$.
Hastatic order is the condensation of this bosonic spinor (cf. Fig. 6)
\begin{equation}\label{}
\hast_{\sigma }\dg\chi_{\alpha} \rightarrow \langle\hat  
\hast_{\sigma }\dg \rangle 
\chi_{\alpha }. 
\end{equation}
This may be viewed as a symmetry-breaking Gutzwiller projection.  
The resulting quadratic Hamiltonian involves a symmetry-breaking 
hybridization between the conduction electrons and the pseudofermions.
Because experimentally the HO and the AFM share a single commensurate wavevector\cite{Villaume08,Hassinger10}
$Q = (0,0, \frac{2\pi}{c})$, we use this wavevector in the description
of the HO state where $<\Psi_\pm > = |\Psi| \exp^{\pm \frac{i(\vec{Q}\cdot\vec{R}_j + \phi)}{2}}$
where the internal angle $\phi$ rotates the hastatic spinor in the basal plane.  Exploiting
the gauge symmetries of the problem, we can simplify the valence-fluctuation Hamiltonian to read
\begin{equation}
H_{VF} = \sum_\bk c\dg_\bk
\mathcal{V}_6(\bk) \chi _\bk + c\dg_\bk \mathcal{V}_7(\bk) \chi _{\bk+\bQ} +
\mathrm{h.c.}     
\end{equation}
where the hybridization form factors are 
$\mathcal{V}_7(\bk) = V_7 \Phi_7{\dg}(\bk) \sigma_1$ and $\mathcal{V}_{6}
(\bk )= V_{6}\Phi_{6}{\dg} (\bk )$ and there is uniform
($\Gamma_6$) and staggered ($\Gamma_7^-$) hybridization in the two channels.

\fg{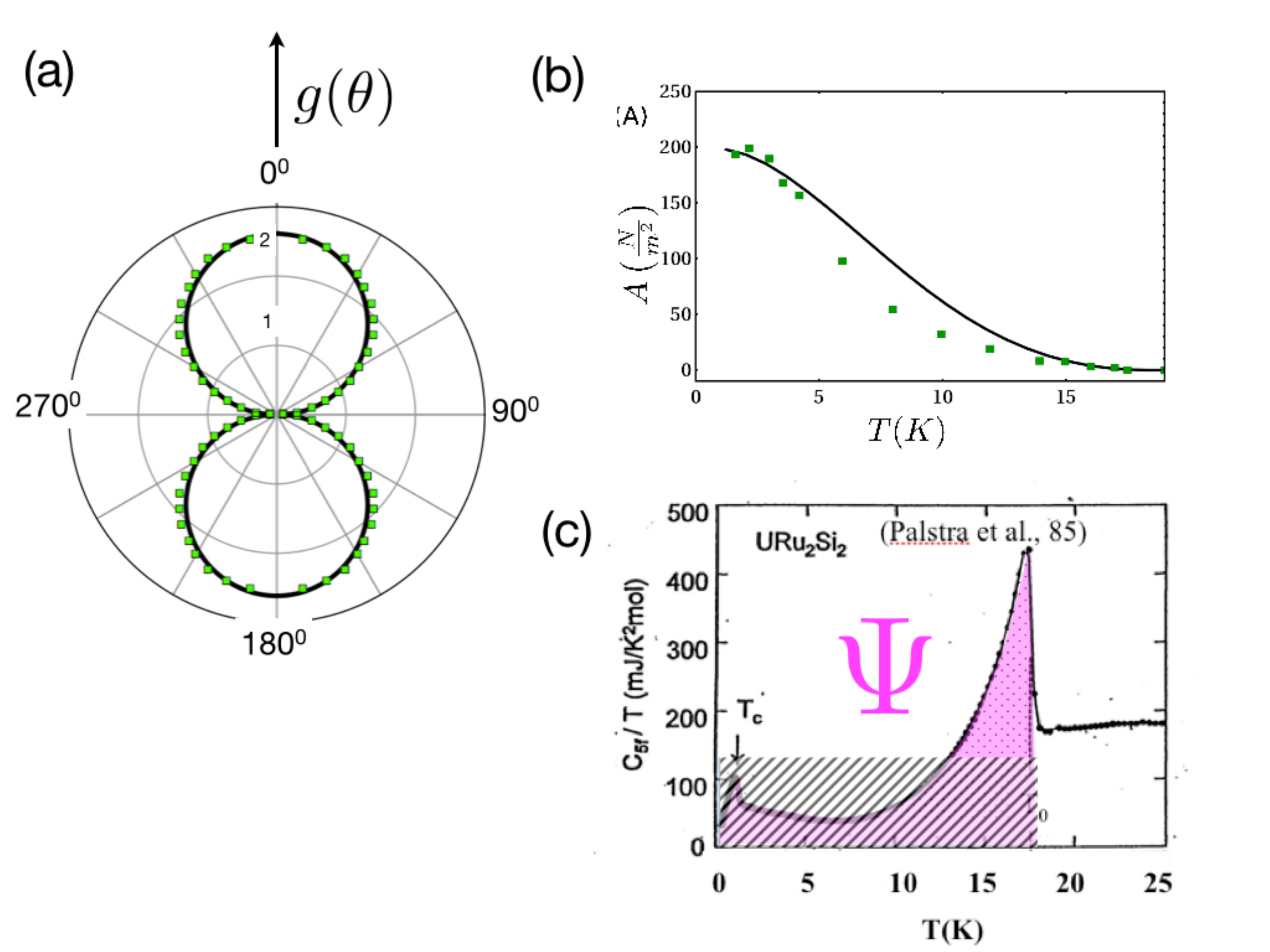}{fig6}{Consistency calculations from the hastatic theory
indicating good agreemnt with experiment for (a) 
the anisotropic g-factor of the quasiparticles  (b) the anistropic susceptibility $\chi_{xy}$ and the (c) entropy associated with the hidden order transition.}

This mean-field hastatic model can be used to calculate observable
quantities, both to check consistency with known measurements and also
to make predictions for future experiment.  The full anistropic
g-factor is a combination of $f$-electron and conduction electron
contributions and the result for the Fermi-surface averaged g-factor
as a function of field-angle to the c-axis is displayed in Fig. 3,
demonstrating good consistency with previous experiment.  Magnetometry
measurements indicate the development of an anisotropic basal-plane
spin susceptibility, $\chi_{xy}$ at the HO transition,\cite{Okazaki11}
and this result is interpreted as resonant scattering off the Ising U
moments and calculated $\chi_{xy}$ within our model; the result
compares well with experiment as displayed in Fig. 7.  The development
of hastatic order in the lattice at the HO transition liberates a
large entropy\cite{Bolech02} of condensation, $\frac{S}{N} \sim
\frac{1}{2} k_{B}\ln 2$ a natural consequence of a Majorana zero-mode
in two-channel Anderson impurity physics.

Having established consistency, we now discuss the resulting predictions.
The gap to longitudinal spin fluctuations in the hastatic
state, and the highly anisotropic nonlinear susceptibility anomaly has
been discussed earlier. The detailed microscopic model can be used to
determine the magnitude of this quantity.  Within the hastatic theory,
there is time-reversal breaking in both the HO and the AFM phases and
there must be some physical manifestation of this phenomenon in the HO
state.  Below $T_{HO}$, this theory predicts a small conduction
electron and f-electron moment in the basal plane. This will be
discussed more when we review recent experiment.  The hastatic theory
also predicts a hybridization gap that breaks tetragonal symmetry below
$T_{HO}$. The resonant scattering via this hybridization  leads to a
resonant nematicity in the local density of states
that is predicted to be a maximum at energies corresponding to the  Kondo
resonance: this signal should be observable 
in STM and ARPES measurements.
 
\fg{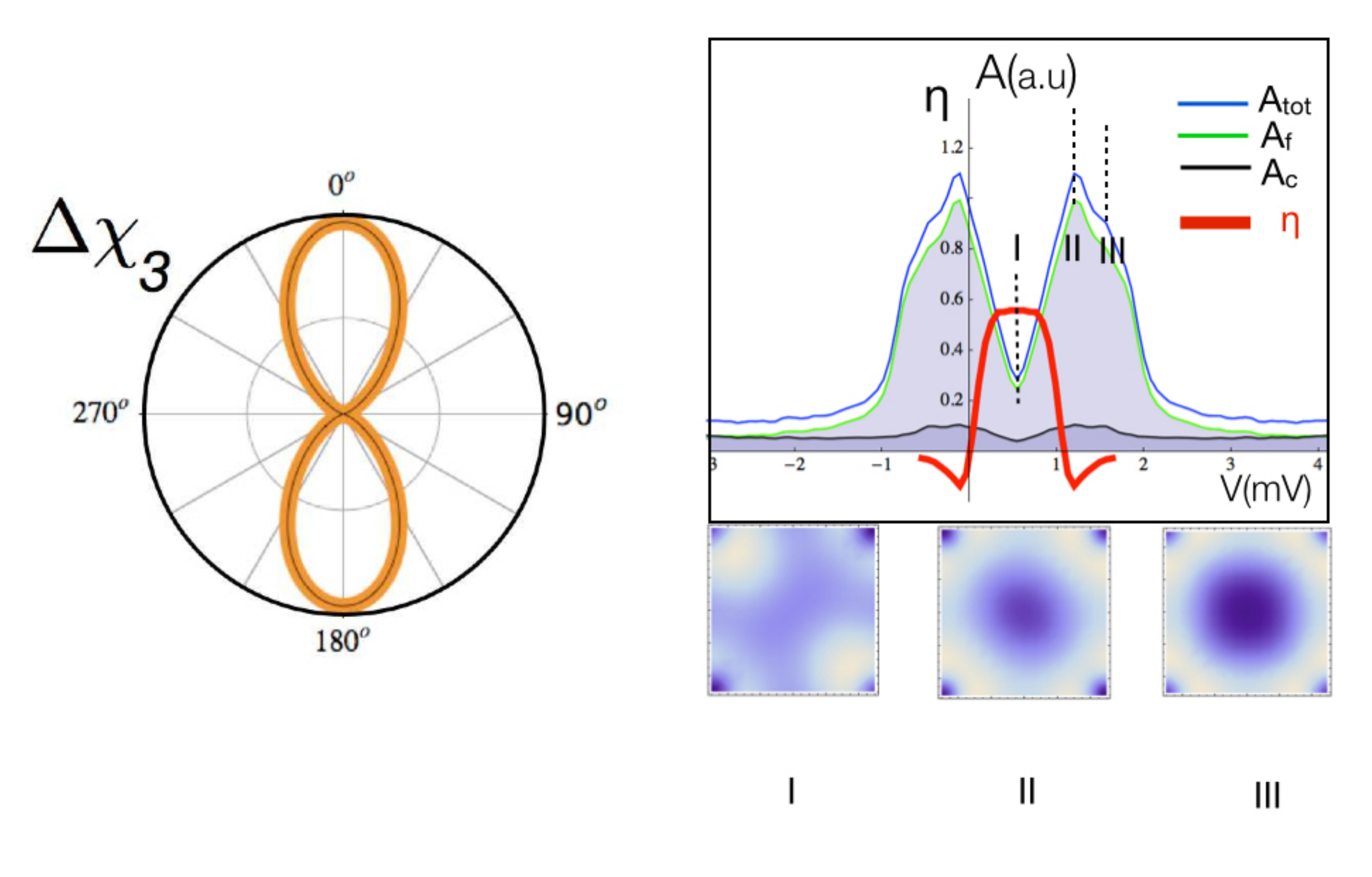}{fig7}{Predictions from the hastatic theory for the (a) anisitropy of the $\chi_3$ anomaly and  the  (b) energy-dependent resonant 
nematicity}

\section{Can Landau order parameters fractionalize?}  

A broader implication of hastatic order is the possibility of a 
new type of Landau order parameter, one that transforms under
double-group (half-integer spin) group representations.
Conventionally Landau theory in electronic systems 
is based on the formation and
condensation of two-body bound-states, described by a Wick contraction
of two electron field operators. 
The resulting order parameter carries an integer spin. 
For example in magnetism, the
development of a magnetic order parameter $\vec{M} (x)$ is given
by the contraction
\begin{equation}\label{}
\contracty{\psi \dg_{\alpha } (x)\psi_{\beta } (x)}= \vec{\sigma}_{\alpha \beta }
\cdot \vec{M } (x)
\end{equation}
By contrast, s-wave superconductivity is based on the formation of
spinless
bosons given by the contraction
\begin{equation}\label{}
\contracty{\psi _{\up} (1)\psi_{\dw} (2)}= -F(1 - 2),
\end{equation}
where $F (1-2 )= -\langle T\psi _{\up} (1)\psi_{\dw} (2)\rangle $ is
the anomalous Gor'kov Greens function
which breaks the gauge system of the underlying system.
The take-home message from conventional two-body
condensation is that when the two-body bound-state wavefunction
carries a quantum number (e.g. charge or spin), a symmetry is broken.
However under this scheme, all order parameters are bosons that carry
integer spin.

Hastatic order carries half-integer spin and cannot develop via this
mechanism. We are then led to the question of whether it is possible
for Landau order parameters to transform under half-integer
representations of the spin rotation group.  At first sight this
impossible for all order parameters are necessarily bosonic, and
bosons carry integer spin.  However the connection between spin and
statistics is strictly a relativistic idea that depends on the full
Poincare invariance of the vacuum.  This invariance is lost in
non-relativistic condensed matter systems suggesting the possibility
of order parameters with half-integer spin that transform under
double-group representations of the rotation group.  Spinor order
parameters involving ``internal'' quantum numbers are well known in
the context of two-component Bose-Einstein condensates.  The Higgs
field of electroweak theory is also a two-component spinor.  However
in neither case does the spinor transform under the physical rotation
group.  Moreover it is not immediately obvious how such bound-states
emerge within fermionic systems.

In the mean-field formulation of hastatic order,\cite{Chandra13} a spin-1/2 order parameter develops as a consequence
of a factorization of a Hubbard operator that connect the Kramers and
non-Kramers states; it is a tensor operator that 
corresponds to the three-body combination 
\begin{equation}\label{}
X_{\alpha \sigma } (R)\equiv
\vert f^{2}\alpha \rangle   
\langle f^{1}\sigma \vert
= \Lambda_{\alpha \sigma }^{abc} (R;1,2,3) \psi\dg _{a } (1)\psi\dg _{b} (2)\psi_{c} (3),
\end{equation}
where we have used 
the short-hand notation $1 \equiv R_{1}$ etc. and 
\begin{equation}\label{}
\Lambda_{\alpha \sigma }^{abc} (R;1,2,3)= 
\langle R_{1},a;R_{2},b\vert \hat X_{\alpha \sigma } (R)\vert R_{3},c\rangle 
\end{equation}
defines the overlap between the Hubbard operators and the bare
electron states. 
In a
simple model, this three body wavefunction is local, $\Lambda_{\alpha \sigma
}^{abc} (R;1,2,3)= \Lambda_{\alpha \sigma }^{abc}\delta
(R-1)\delta (R-2)\delta (R-3)$.
The factorization of the Hubbard operator into 
a spin-1
fermion and a spin-1/2 boson 
\begin{widetext}
\begin{equation}\label{}
X_{\alpha \sigma } (R)\rightarrow 
\chi\dg _\alpha (R)
\left\langle 
 \Psi_{\sigma } (R)\right\rangle 
,
\end{equation}
then represents a ``fractionalization'' of the three body operator. 
Written in terms of the microscopic electron fields, this
becomes 
\begin{eqnarray}\label{l}
\Lambda_{\alpha \sigma }^{abc} (R;1,2,3)\contract{\psi\dg _{a} (1)\psi\dg_{b} (2)\psi _{c} (3)}&=&
\chi\dg  _{\alpha } (R)
\left\langle \phantom{\sum}\hskip -5mm \Psi_{\sigma }(R) \right\rangle
.\cr&&
\end{eqnarray}
This expression can be inverted to give the three body contraction
\begin{eqnarray}\label{l}
\contract{\psi\dg _{a} (1)\psi\dg _{b} (2)\psi _{c} (3)}&=&
\sum_{R}G_{abc}^{\alpha \sigma}(1,2,3;R) 
\chi\dg  _{\alpha } (R)
\left\langle \phantom{\sum}\hskip -5mm \Psi_{\sigma }(R) \right\rangle
,\cr&&
\end{eqnarray}
where 
$G_{abc}^{\sigma\alpha}(1,2,3;R) =[ \Lambda_{ \sigma\alpha  }^{abc} (R;1,2,3)]^{*}$.
The asymmetric decomposition 
of a three-body Fermion state into a binary combination of
boson and fermion is a  fractionalization process. If the boson in
binary carries a quantum number, when it condenses we have the
phenomenon of 
{\sl ``order parameter fractionalization''}. \\

\end{widetext}
Fractionalization is well established for
excitations of low dimensional systems, such the one dimensional
Heisenberg spin chain and the fractional quantum 
Hall effect.\cite{Jackiw76,Su79,Laughlin99,Castelnuovo12}
The hastatic ordering process involves the order parameter
fractionalization into binary combination of a
condensed half-integer spin boson and an integer spin fermion.
Unlike pair or exciton
condensation, the order parameters formed by this mechanism transform
under double group representations of the underlying symmetry groups,
and thus represent a fundamentally new class of broken symmetries. 
We look forward to investigating this ``order parameter fractionalization'' well
beyond the realm of \urs.  The proposed three-body bound-state has a nonlocal
order parameter, and it may be possible to identify a dual theory with
a local order parameter that breaks a global symmetry.
\fg{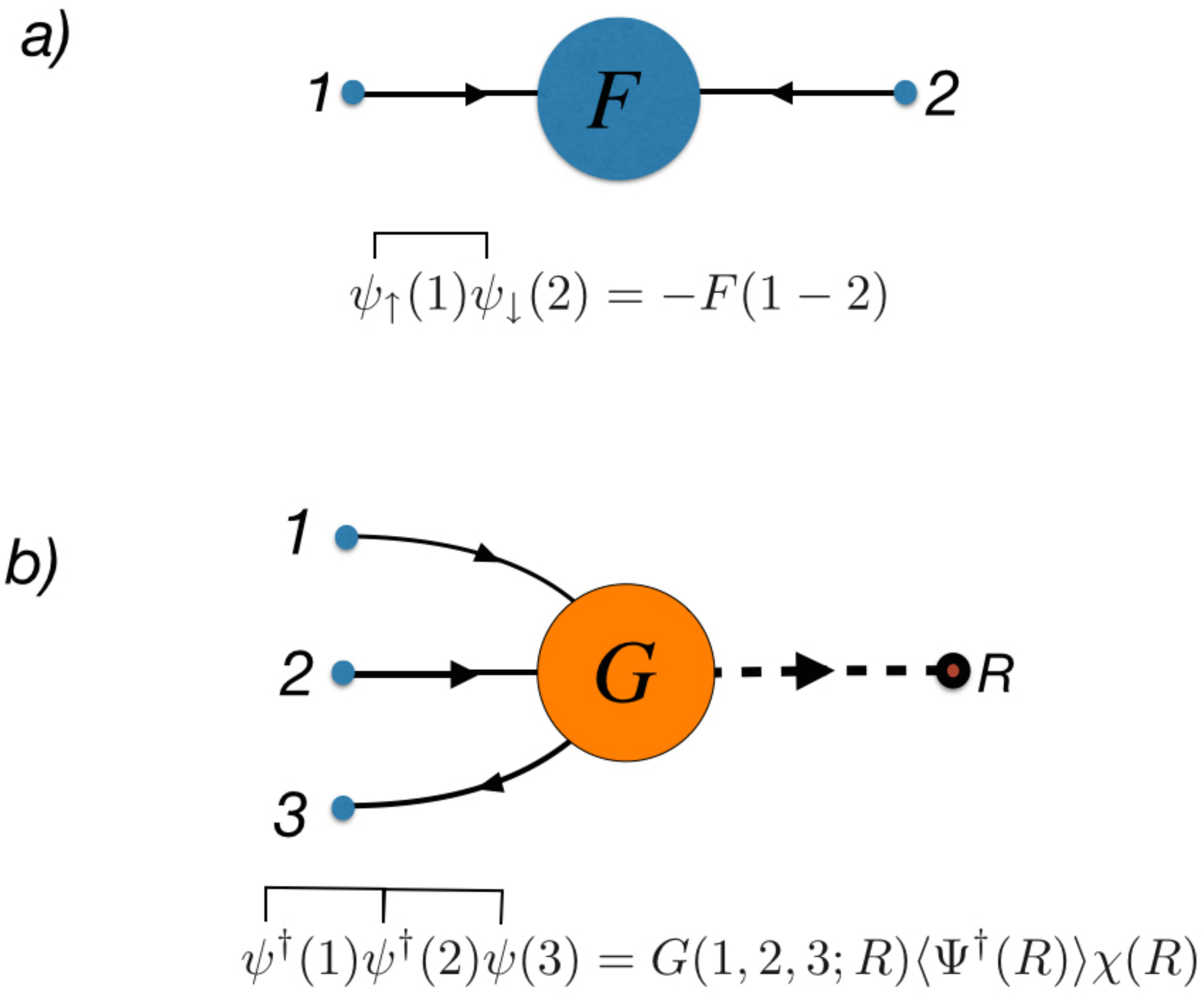}{fig8}{Schematic Feymann diagrams indicating (a) two-body  (b) and three-body electronic bound-states where in the latter case spin indices have been suppressed for pedagogical simplicity.}

\section{Discussion of Recent Experiments...With Specific Requests} 

Let us now return to the situation in \urs.  We mentioned earlier
that hastatic order leads to a prediction of a basal-plane moment of
order $T_{K}/D$\cite{Chandra13,Flint14}, where $T_{K}$ and $D$ are the
Kondo temperature and band-width respectively. 
The tranverse moment in our mean-field treatment includes both
conduction and f-electron contributions
which point in perpendicular directions. The ratio $T_{K}/D$ is very
sensitive to the degree of mixed valence of the $5f^{2}$ state. 
Our original calculation assumed a $20\%$ mixed valence,
leading to a basal plane moment of order $0.01 \mu_B$. 
Recent high-resolution experiments\cite{Das13,Metoki13,Ross14} 
have failed to observe a
transverse moment of this magnitude, and have placed a bound 
$\mu_{\perp }< 0.0011\mu_{B}$ on the ordered
transverse moment of the uranium ions. Paradoxically
various other probes including 
X-rays, $\mu$-spin resonance and
NMR\cite{Caciuffo14,Amitsuka03,Bernal04,Takagi12} have detected the
presence of static basal moments 
on the order of $0.005 \mu_B$ that would be consistent
with a more integral valent scenario for the $U$ ions. 

%


These remaining ambiguities suggest 
we need to reconsider 
the calculation of the transverse moment and
understand why it is so small if not absent.
There are a number of interesting 
possibilities:
\begin{itemize}
\item {\sl Fluctuations}. 
The hastatic theory, in its current version, ignores
fluctuations of the spinor order that 
will reduce the transverse moment.
Gaussian fluctuations of the corresponding Schwinger boson
field  are needed to describe the development of the incoherent Fermi
liquid observed to develop at 
$T>T_{HO}$ in optical, tunneling and thermodynamic measurements 
.\cite{Schmidt10,Aynajian10,Park11,Haraldsen11}

\item {\sl Uranium Valence}. As mentioned already, the predicted transverse moment
is sensitive to the $5$f valence, and would  be much
reduced by a vicinity to integral valence.  Moreover, it should be proportional only to the \emph{change} in valence
between $T_{HO}$ and the measurement temperature, which will be signifcantly smaller than the high-temperature mixed valency.
It would be very helpful to have low temperature 
probes of the 5f-valence. 

\item {\sl Domain Size}. The X-ray,\cite{Caciuffo14} 
muon,\cite{Amitsuka03} torque
magnetometry\cite{Okazaki11} and NMR measurements\cite{Bernal04,Takagi12} 
that indicate either
a static moment or broken tetragonal symmetry are all carried out on
small samples, whereas the neutron measurements involve large
ones.\cite{Das13,Metoki13,Ross14} The discrepancy between the two classes of measurement may indicate the formation of small  hidden order
domains. Such domain structure might be the result of 
random pinning\cite{Imry75} of the transverse moment 
by defects of random strain fields. 
The situation in \urs is somewhat analogous
to that in Sr$_{2}$RuO$_{4}$, where there is evidence for broken
time-reversal symmetry breaking with a measured Kerr effect
and $\mu$SR to support chiral p-wave superconductivity, but
no surface currents have yet been observed.\cite{Kallin12}  Domains are
an issue in this system too.

\item {\sl Continuous versus discrete order}. The current mean-field theory
has the transverse hastatic vector $\Psi \dg \vec{\sigma }\Psi $
pointing in one of four possible directions at each site,
corresponding to a four-state clock model.  The
tunneling barrier between these configurations is very small, leaving
open the possibility that at long distances the residual physics is
that of an xy order parameter. Such xy order would then give rise to
a kind of spin-superfluid, in which the persistent spin currents
avoid the formation of a well-defined static staggered moment. 

\end{itemize}  

There are a number of important measurements that would help to
resolve some of the current uncertainties and test some of the
outstanding predictions:

\begin{enumerate}

\item {\sl Giant Anisotropy in $\Delta \chi_3\propto \cos^4 \theta $}.
This measurement is important to confirm that that the Ising
quasiparticles are associated with the development of the hidden
order.
      
\item {\sl dHvA on all the heavy Fermi surface pockets}. We expect that the 
      heavy quasiparticles on the $\alpha $ $\beta $ and $\gamma$
      orbits will all exhibit the multiple spin zeros of Ising
      quasiparticles. At present, only the $\alpha$ orbits have been
      measured as a function of field orientation. 

\item {\sl Spin zeros in the AFM phase? (Finite pressure)}
      If the AFM is also hastatic, then we expect the spin zeros to
      persist into the finite pressure AFM pahse. 

\end{enumerate}



\section{The Challenges Ahead} 

The observation of Ising quasiparticles in the hidden order state
\cite{Brison95,Okhuni99,Altarawneh11,Altarawneh12}
represents a major challege to our understanding of \urs; to our
knowledge this is only example of such anisotropic mobile electrons.
It plays a central role in the hastatic proposal, and a key
question is whether this phenomenon can be accounted for
in other HO theories:

\begin{enumerate}  

\item {\sl Can band theory account for the quasiparticle Ising
anisotropy observed in \urs?} Recent advances in the understanding
of orbital magnetization\cite{Xiao05,Thonhouser05,Xiao06} make it possible
to compute the g-factor associated with conventional Bloch waves.  It
would be particularly interesting to compare the quasiparticle $g
(\theta )$ computed in a density functional treatment of \urs with
that observed experimentally.

\item {\sl Can other  $5f^2$ theories account for the multiple spin
zeroes and the upper bound $\Delta  < 1K$ 
on the spin degeneracy of the heavy fermion bands? }
In particular, is it possible to account for the observed spin 
zeros without invoking a non-Kramers $5f^{2}$ doublet?

\end{enumerate}

In summary, any theory of hidden order has to be
able to explain the giant Ising quasiparticle anisotropy in
 \urs.  The smooth pressure-dependence of the Fermi surfaces
between the Hidden Order and the Antiferromagnetic states is also
mysterious;\cite{Hassinger10} it's as if the differences between the two order parameters are ``invisible'' to the two Fermi surfaces!  Finally there is the 
key question of why superconductivity only emerges from the hidden order
state.

We have benefitted from inspiring discussions with our colleagues
who include C. Batista, C. Broholm, K. Haule, N. Harrison, G. Kotliar,
G. Lonzarich, J. Mydosh, K. Ross and J. Schmalian. 
PC and PC are grateful to Trinity College, Cambridge  and the Cavendish
Laboratory where this article was completed.   This work
was supported by the National Science Foundation grants NSF-DMR-1334428
(P. Chandra) and DMR-1309929 (P. Coleman), and by the Simons Foundation (R. Flint).

\end{document}